\documentclass[aps,prb,reprint,superscriptaddress,longbibliography]{revtex4-2}
\usepackage{amsmath,amssymb,bm}
\usepackage{graphicx}
\usepackage{physics}
\usepackage{hyperref}
\usepackage{xcolor}
\hypersetup{colorlinks=true,linkcolor=blue,citecolor=blue,urlcolor=blue}
\newcommand{\hc}{\mathrm{H.c.}}

\begin{document}

\title{Dipolar Exchange and Emergent Electric-Dipole-Carrying Quasiparticles in Electronic Ferroelectrics}

\author{Ping Tang}
\email{tang.ping.a2@tohoku.ac.jp}
\affiliation{Institute for Materials Research, Tohoku University, Sendai 980-8577, Japan}
\author{Eiji Saitoh}
\affiliation{Department of Applied Physics, The University of Tokyo, Tokyo 113-8656, Japan.}
\date{\today}

\begin{abstract}
Electronic correlations underpin diverse spontaneously ordered states in matter, whose collective excitations emerge as quasiparticles that encode the underlying order. While spin exchange interactions give rise to magnetic order and its quasiparticle excitations known as magnons, their counterparts in electronic ferroelectrics remain poorly understood. Here, we derive an analogous quantum dipolar-pseudospin exchange model for electronic ferroelectricity and identify the associated bosonic quasiparticles that carry intrinsic electric dipole moments. Unlike ferrons in conventional lattice-based ferroelectrics, these dipolar quasiparticles
arise entirely from collective electronic
dynamics without involving lattice vibrations. Their excitation spectrum exhibits a linear Stark shift under a static electric field, while coupling to an oscillating electric field enables their coherent excitation, leading to colossal, resonantly enhanced second-harmonic generation and optical rectification in the terahertz regime. These dipolar quasiparticles offer promising opportunities for electrically tunable nonlinear optics and open a route toward extending the paradigm of transport in electronic systems beyond conventional charge currents to electric dipole currents.
\end{abstract}

\maketitle

\emph{Introduction.---}Electronic correlations underpin diverse spontaneously
ordered states in condensed matter~\cite{Imada1998,Dagotto2005}, including magnetic,
orbital, and charge order, as well as superconductivity. Their collective excitations can be quantized as emergent quasiparticles that encode the underlying order and exhibit particle-like behavior~\cite{PhysRev.127.965}. A paradigmatic example is magnetic order
arising from exchange interactions, with its
elementary excitations known as magnons. As carriers of spin angular momentum, magnons
enable spin transport without accompanying
charge currents~\cite{Kajiwara2010,cornelissen2015long},
making them attractive for low-dissipation
spintronic applications~\cite{kruglyak2010magnonics,Chumak2015}. Likewise, orbital-ordered states support collective orbital-wave excitations or orbitons~\cite{Ishihara2000,Saitoh2001,vanDenBrink2001,Schlappa2012},
which constitute the orbital counterparts of magnons. More recently, charged bosonic quasiparticles termed bichargons have been predicted to emerge
as collective excitations of a charge-ordered state and mediate charge transport in electrically
insulating systems~\cite{tang2026collective,tang2026bichargon}.

Ferroelectricity represents another fundamental order in matter that shares close phenomenological analogies with magnetism~\cite{spaldin2007analogies}. Its electrically controllable spontaneous polarization enables diverse functionalities, including nonvolatile memories, sensors, and actuators~\cite{xu2013ferroelectric}. In contrast to magnetism, ferroelectricity is typically caused by inversion-symmetry-breaking lattice distortions associated with the condensation
of a soft phonon mode below a critical
temperature~\cite{Cochran1959,Cochran1960}. Ferrons are the collective excitations of such a soft-phonon-condensed polar state that carry intrinsic electric dipoles~\cite{Tang2022Ferron,bauer2022magnonics,Tang2024Ferron,tang2026multi}. Their ability to transport electric dipole moments and heat~\cite{wooten2023electric,tang2023nonlocal,shen2025observation,itoh2026observation}, together with their potential for giant terahertz emission~\cite{Choe2026Ferron,Zhang2026Ferron,jana2026ferron}, opens avenues for novel thermoelectric functionalities based on heat-to-electric-dipole-current conversion~\cite{PhysRevLett.126.187603,PhysRevLett.128.047601,PhysRevApplied.20.050501} and field-controlled terahertz optical applications~\cite{Choe2026Ferron,Zhang2026Ferron,jana2026ferron}. Ferroelectricity, however, need not originate from structural distortions but can instead emerge from spontaneous polar ordering of correlated electrons~\cite{Portengen1996,vanDenBrink2008}, in close microscopic analogy to magnetic order. Although several candidate electronic ferroelectrics have been identified or proposed~\cite{Monceau2001,ikeda2005ferroelectricity,Yamauchi2009,Lunkenheimer2012}, the nature of their quasiparticle
excitations remain poorly understood~\cite{naka2012collective,PhysRevLett.110.106401,
ishihara2014electronic,Adachi2023}. Unlike ferrons in conventional structural ferroelectrics, which are rooted in anharmonic lattice dynamics~\cite{Tang2022Ferron,tang2026multi}, collective excitations of electronic ferroelectric order involve fundamentally different degrees of freedom associated with correlated electron dynamics. Whether such electronic collective excitations can give rise to purely electronic ferron-like quasiparticles capable of transporting electric dipole moments, and how these quasiparticles respond to external fields, remain open questions.

In this Letter, we derive an effective quantum dipolar-pseudospin exchange model for electronic ferroelectricity, in close analogy to the Heisenberg spin-exchange model of magnetic order. By quantizing the collective dipolar-pseudospin-wave excitations, we identify the associated bosonic quasiparticles that carry intrinsic electric dipole moments and enable electric dipole transport even in electrically insulating systems. Unlike conventional lattice-based ferrons, these dipolar quasiparticles are purely electronic in origin, arising from collective fluctuations of the electronic charge distribution without involving lattice vibrations. We further show that their excitation spectrum can be tuned by a static electric field through a linear Stark effect, while their coupling to alternating electric fields generates nonlinear polarization responses, including exceptionally large, resonantly enhanced second-harmonic generation and optical rectification in the terahertz regime. These characteristic responses provide spectroscopic signatures of the dipolar quasiparticles and open avenues for electrically tunable nonlinear optical functionalities.

\begin{figure}
    \centering
    \includegraphics[width=8.2 cm]{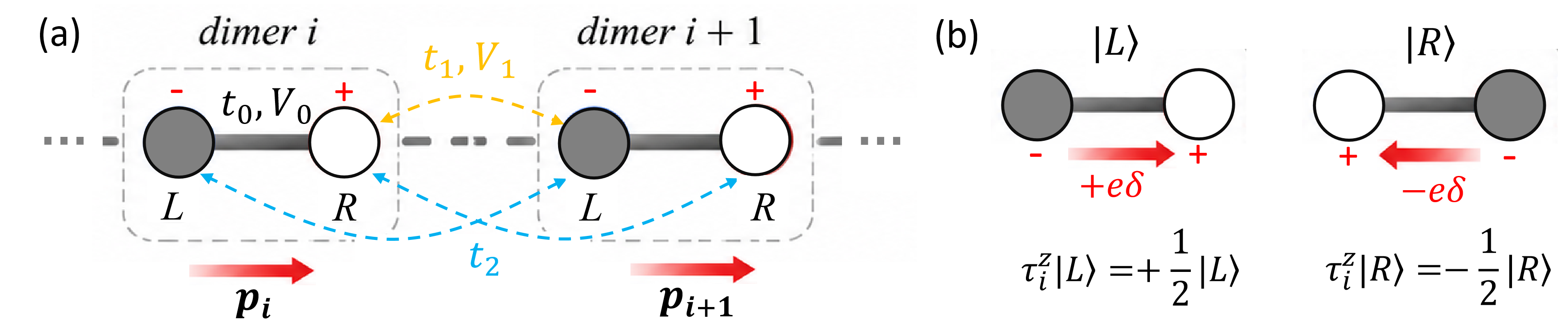}
    \caption{(a) Illustration of a dimerized one-dimensional extended Hubbard model for electronic ferroelectricity with one electron per dimer. Ferroelectricity emerges from the spontaneous collective ordering of electrons occupying the same side (shaded) of each dimer, while the underlying lattice remains centrosymmetric. (b) Two intradimer polar states $\vert L\rangle$ and $\vert R\rangle$, corresponding to left- and right-side electron occupation and eigenvalues $\pm 1/2$ of the longitudinal pseudospin operator $\tau_i^z$, respectively.}
    \label{fig-1}
\end{figure}

\emph{Dipolar-pseudospin exchange.---}We consider an extended Hubbard model for electronic ferroelectricity on a dimerized one-dimensional chain with an average of one electron per dimer, where each dimer $i$ consists of two sites labeled $L$ and $R$, as illustrated in Fig.~\ref{fig-1}. The microscopic Hamiltonian reads
\begin{align}
H=&-t_{0}\sum_i(c_{iL}^{\dagger}c_{iR}
+\hc)-t_{1}\sum_{i}(c_{iR}^{\dagger}c_{i+1L}+\hc)\nonumber\\
&-t_{2}\sum_{i\alpha}(c_{i\alpha}^{\dagger}c_{i+1\alpha}
+\hc)+V_{0}\sum_i n_{iL}n_{iR} \nonumber\\&+V_{1}\sum_{i}n_{i+1L}n_{iR},
\label{eq:micro}
\end{align}
where $c_{i\alpha}^{\dagger}$ creates an electron on site $\alpha=L,R$ of dimer $i$, and $n_{i\alpha}=c_{i\alpha}^{\dagger}c_{i\alpha}$ is the corresponding number operator. Here $t_0$ denotes the intradimer hopping, while $t_1$ and $t_2$ describe the interdimer
hoppings between adjacent and equivalent sites
of neighboring dimers, respectively. $V_0>0$ and $V_1>0$ represent the intradimer and interdimer Coulomb repulsions between nearest-neighbor sites, respectively, with longer-range Coulomb interactions neglected. An unequal electronic occupation of the two
intradimer sites generates a local electric dipole, while the collective alignment
of these dipoles gives rise to spontaneous
electronic polarization. For our purposes, we disregard the spin degree of freedom by considering spinless fermions, for which double occupation of a single site is forbidden by $(c_{i\alpha}^{\dagger})^2=0$. This minimal spinless model may be viewed as a simplified
description of the charge sector of strongly
correlated electronic systems in the limit of
strong on-site Hubbard repulsion, where doubly
occupied sites are projected out. It captures a generic mechanism of electronic ferroelectricity
relevant to organic charge-transfer salts,
including (TMTTF)$_2$X
~\cite{PhysRevLett.85.1698,Monceau2001,Giovannetti2015}
and dimer-Mott systems based on
$\kappa$-(BEDT-TTF)$_2$X
~\cite{Hotta2010,naka2010electronic,Lunkenheimer2012},
as well as mixed-valence transition-metal oxides,
such as Fe$_3$O$_4$~\cite{Yamauchi2009}.

We focus on the strong-coupling dimer regime, $V_0\gg V_{1}, |t_n|$, where configurations involving an empty dimer and a doubly occupied dimer are energetically costly due to the large intradimer Coulomb repulsion $V_0$. The low-energy sector is then restricted to configurations with one electron per dimer,
\begin{equation}
n_{iL}+n_{iR}=1,
\label{num}
\end{equation}
where the fermion can occupy either the left site, $|L\rangle_i=c_{iL}^{\dagger}|0\rangle$, or the right site, $|R\rangle_i=c_{iR}^{\dagger}|0\rangle$. These two local states define a
pseudospin-$1/2$ degree of freedom on each dimer through
\begin{equation}
\tau_i^z=\frac{1}{2}(n_{iL}-n_{iR}),\quad
\tau_i^+=c_{iL}^{\dagger}c_{iR},\quad
\tau_i^-=c_{iR}^{\dagger}c_{iL},\label{psed}
\end{equation}
where the longitudinal component $\tau_i^z$ describes the charge imbalance between the two sites within a dimer, distinguishing $|L\rangle_i$ and $|R\rangle_i$ with eigenvalues $+1/2$ and $-1/2$, respectively, as illustrated in Fig.~\ref{fig-1}. The electric dipole moment associated with each dimer is 
\begin{equation}
p_i=-e\left(-\frac{\delta}{2}n_{iL}
+\frac{\delta}{2}n_{iR}\right)
=d_{0}\tau_i^z,
\label{eq}
\end{equation}
where $e>0$ is the elementary charge and $d_{0}=e\delta$ is the unit of the intradimer dipole moment, with $\delta$ the separation between the two sites
within a dimer. In contrast, the ladder operators $\tau_i^\pm$ induce transitions between the two local states, $|L\rangle_i=\tau_i^+|R\rangle_i$ and $|R\rangle_i=\tau_i^-|L\rangle_i$, corresponding to the reversal of the intradimer electric dipole moment.

The interdimer hopping terms couple the low-energy
manifold defined by Eq.~(\ref{num}) to high-energy
virtual states containing an empty dimer and a
neighboring doubly occupied dimer. As derived in the Supplemental Material~\cite{SM},
a second-order Schrieffer--Wolff transformation
yields the effective pseudospin Hamiltonian
\begin{align}
H_{\rm eff}\simeq\sum_i\left[J_{z}\tau_i^z\tau_{i+1}^z+
\frac{J_{\perp}}{2}\left(\tau_i^+\tau_{i+1}^-+\text{H.c.}\right)-h_x\tau_i^x \right],
\label{eq:XXZ}
\end{align}
where
$J_z\simeq -V_1+(4t_2^2-2t_1^2)/V_0$
and $J_\perp\simeq 4t_2^2/V_0>0$
denote the longitudinal and transverse
dipolar-pseudospin exchange couplings between
neighboring dimers, respectively, while
$h_x=2t_0$ is the uniform transverse field
arising from intradimer hopping.
The interdimer hopping $t_1$ and Coulomb
repulsion $V_1$ favor ferromagnetic Ising
coupling, whereas $t_2$ generates an isotropic
antiferromagnetic Heisenberg exchange in
pseudospin space, analogous to superexchange mediated by
virtual hopping in conventional Mott insulators~\cite{Anderson1959,Takahashi1977,MacDonald1988}. Mixed $t_1$-$t_2$ virtual processes do \emph{not} contribute to the effective nearest-neighbor pseudospin interactions at second order, since two successive hops involving different
interdimer channels cannot restore the
one-electron-per-dimer configuration. Unlike the staggered pseudospin order representation of charge order in the half-filled Hubbard model~\cite{tang2026pseudospin}, the present anisotropic Heisenberg pseudospin model supports ferroelectricity characterized by uniform longitudinal ordering, $\langle\tau_i^z\rangle\neq0$, corresponding to the collective alignment of intradimer electric dipoles, closely analogous to that described by the transverse-field Ising model of order-disorder ferroelectrics~\cite{de1963collective,blinc1972dynamics,Tang2024Ferron}.

Eq.~(\ref{eq:XXZ}) supports distinct
ground-state phases governed by the competition
among $J_z$, $J_\perp$, and $h_x$.
In the absence of a transverse field ($h_x=0$),
the model reduces to the conventional spin-$1/2$
XXZ chain.
For $J_z<-J_\perp$, the ground state exhibits
ferromagnetic Ising pseudospin order,
$\langle\tau_i^z\rangle=\pm\tau_0$,
corresponding to an inversion-symmetry-breaking
ferroelectric state.
In the intermediate regime
$-J_\perp<J_z<J_\perp$, the ground state is
a gapless Luttinger liquid~\cite{giamarchi2004quantum},
characterized by quasi-long-range transverse
pseudospin correlations and the absence of
long-range longitudinal order.
For $J_z>J_\perp$, the ground state instead
exhibits antiferromagnetic Ising pseudospin order,
$\langle\tau_i^z\rangle=(-1)^i\tau_0$,
corresponding to an antiferroelectric state
with alternating intradimer charge configurations.

A finite transverse field $h_x$ induces quantum
tunneling between the left- and right-site
charge configurations, tending to align the
pseudospins along the $x$ direction.
Starting from either longitudinally ordered
Ising phase, a weak transverse field preserves
the corresponding longitudinal order, while
increasing $h_x$ progressively suppresses it.
At a critical field $h_x=h_c$, the longitudinal
order parameter vanishes, signaling a quantum
phase transition to a nonpolar phase.
For sufficiently strong $h_x$, the ground state
becomes predominantly polarized along the
$x$ direction, with $\langle\tau_i^z\rangle=0$, corresponding to coherent superpositions of
the left- and right-site charge configurations.

\emph{Dipolar quasiparticle excitations.---}We focus on the ferroelectric regime, $J_z<-J_\perp$ and $h_x<h_c$, which hosts
two inversion-related ferroelectric ground states
at zero temperature. In strictly one-dimensional systems with short-range
interactions, thermally excited domain walls destroy
long-range Ising order at any finite temperature~\cite{Ising1925}.
In realistic materials, however, interchain couplings
can stabilize long-range ferroelectric order by suppressing
the proliferation of thermally excited domain walls~\cite{Scalapino1975}.
The idealized one-dimensional model considered here
nevertheless captures the essential physics of collective quasiparticle excitations in electronically driven ferroelectrics
without explicitly incorporating interchain couplings.

Assuming spatially uniform order, we parameterize the pseudospin expectation value
as $\langle\boldsymbol{\tau}_i\rangle
=\tau_0(\sin\theta,0,\cos\theta)$,
where $\tau_0$ denotes its magnitude and
$\theta$ is the canting angle relative to
the $z$ axis. Minimizing the classical ground-state energy
of Eq.~(\ref{eq:XXZ}) with respect to $\theta$
yields
$\sin\theta=h_x/h_c$, where $h_c=2\tau_0(J_\perp-J_z)$ is the
mean-field critical transverse field beyond
which the ferroelectric order vanishes. To describe small fluctuations about the canted classical ground state, we introduce
a local coordinate frame $(\tilde{x},\tilde{y},\tilde{z})$,
with the $\tilde{z}$ axis aligned with $\langle\boldsymbol{\tau}\rangle$. The pseudospin operators in the laboratory and rotated frames are related by $\tau_i^x=\cos\theta\,\tilde{\tau}_i^x+\sin\theta\,\tilde{\tau}_i^z$, $\tau_i^y=\tilde{\tau}_i^y$, and $\tau_i^z=-\sin\theta\,\tilde{\tau}_i^x+\cos\theta\,\tilde{\tau}_i^z$. Small fluctuations of the rotated pseudospins are
described by the Holstein--Primakoff transformation
$\tilde{\tau}_i^z=\tau_0-a_i^\dagger a_i$,
$\tilde{\tau}_i^+=(2\tau_0-a_i^\dagger a_i)^{1/2}a_i$,
and
$\tilde{\tau}_i^-=a_i^\dagger
(2\tau_0-a_i^\dagger a_i)^{1/2}$,
where $a_i^\dagger$ ($a_i$) creates (annihilates)
a local bosonic excitation at dimer $i$. For $\theta=0$, such an excitation corresponds
to a local reversal of the intradimer charge
configuration from the left- to the right-occupied state. Expanding the pseudospin Hamiltonian in Eq.~(\ref{eq:XXZ}) to quadratic order in the bosonic operators and Fourier transforming, we obtain
\begin{align}
H_{\rm eff}
=&
E_{\rm cl}
+
\sum_k
\left[
A_k a_k^\dagger a_k
+
\frac{B_k}{2}
\left(
a_k a_{-k}
+a_k^\dagger a_{-k}^\dagger
\right)
\right]\nonumber\\
=&\mathrm{const.}+\sum_k\hbar\omega_k b_k^\dagger b_k
,
\end{align}
where $A_k=-2\tau_{0}J_z+\tau_{0}\left[
J_\perp(1+\cos^2\theta)
+J_z\sin^2\theta
\right]\cos(ka)$ and $
B_k=\tau_0(J_z-J_\perp)
\sin^2\theta\,\cos(ka)$, with $a$ the unit-cell length and $E_{\rm cl}$ the
classical ground-state energy. The terms linear in the bosonic operators vanish
upon imposing the classical ground-state equilibrium condition. The quadratic Hamiltonian is diagonalized by the
Bogoliubov transformation
$a_k=u_kb_k+v_kb_{-k}^\dagger$,
where $u_k^2-v_k^2=1$ and
$u_k^2+v_k^2=A_k/(\hbar\omega_k)$,
yielding the quasiparticle dispersion
$\hbar\omega_k=\sqrt{A_k^2-B_k^2}$. Here, $b_k^\dagger$ and $b_k$ are the corresponding
creation and annihilation operators for quasiparticles
with momentum $k$, respectively.

\emph{Intrinsic dipole moment.---}Unlike the bichargons predicted in the half-filled extended Hubbard model~\cite{tang2026collective,tang2026bichargon}, these bosonic excitations are electrically neutral, as they involve collective fluctuations of the intradimer charge distribution while preserving exactly one fermion per dimer. Nevertheless, these quasiparticles carry intrinsic
electric dipole moments, analogous to magnons
that carry magnetic moments~\cite{rezende2020fundamentals}. For a single quasiparticle
$|k\rangle=b_k^\dagger|\mathrm{vac}\rangle$,
the carried electric dipole moment is
\begin{equation}
p_k
\equiv
\langle k|\mathcal P|k\rangle
-
\langle\mathrm{vac}|\mathcal P|\mathrm{vac}\rangle
=
-d_{0}\cos\theta
\left(u_k^2+v_k^2\right),
\label{eq:quasiparticle_dipole}
\end{equation}
where $\mathcal P=
d_{0}\sum_i\tau_i^z$ is the total
electric dipole operator, and the subtraction removes the zero-point contribution from the quasiparticle vacuum $|\mathrm{vac}\rangle$. $p_k<0$ indicates that the quasiparticle dipole is oriented opposite to the ground-state polarization, reflecting
a reduction in the macroscopic polarization
upon quasiparticle excitation. For $h_{x}=0$, we have $\theta=0$, $u_k=1$, and $v_k=0$, such that each quasiparticle carries one quantum of the elementary electric dipole moment $p_k=-d_{0}$. The quantization of the quasiparticle dipole follows from the $U(1)$ symmetry of Eq.~(\ref{eq:XXZ}) at $h_x=0$, which ensures conservation of the total
$\tau^z$ and hence the total electric
dipole moment, analogous to magnons in uniaxial
(anti)ferromagnets carrying quantized spin
angular momentum of magnitude~$\hbar$~\cite{rezende2020fundamentals,rezende2019introduction}. A finite $h_x$ arising from intradimer hopping, however, breaks the $U(1)$ symmetry and the associated
dipole conservation, rendering the
quasiparticle dipole moment nonquantized
and momentum dependent through the Bogoliubov transformation factor. 
Despite their fundamentally different microscopic origins,
these purely electronic dipolar quasiparticles can,
like ferrons in conventional lattice-based
ferroelectrics~\cite{Tang2022Ferron,bauer2022magnonics,Tang2024Ferron,tang2026multi}, mediate electric dipole responses,
potentially giving rise to a variety of unconventional
transport and dynamical phenomena. 

\emph{Response to external fields.---}We investigate here the response of these dipolar
quasiparticles to static and alternating electric
fields, providing an experimental avenue for their identification and coherent excitation. An external electric field $E$ applied along the
dimer axis couples directly to the intradimer
electric dipoles through
\begin{align}
H_{\rm ext}
=&-Ed_0\sum_i\tau_i^z\nonumber\\
&\simeq gE(b_0+b_0^\dagger)+E\sum_k\big[
-p_kb_k^\dagger b_k\nonumber\\&
+\lambda_k(b_k^\dagger b_{-k}^\dagger
+b_k b_{-k})\big]+\cdots,
\label{eq:stark_bosonic}
\end{align}
which acts as a longitudinal pseudospin
field $h_z=d_0E$. Here, we have expanded $\tau_i^z$ to quadratic
order in the Bogoliubov quasiparticle operators,
with
$g=d_0\sin\theta\sqrt{N_d\tau_0/2}(u_0+v_0)$
and
$\lambda_k=d_0\cos\theta\,u_kv_k$ the single-quasiparticle and
pair-coupling coefficients, respectively,
where $N_d$ is the number of dimers. The linear term drives coherent excitations
of the zero-momentum dipolar mode,
whereas the two quadratic terms describe
the Stark shifts of quasiparticle energies
and the creation or annihilation of
quasiparticle pairs with opposite momenta,
respectively. These couplings give rise to
linear and second-order nonlinear polarization
responses, including resonantly enhanced
second-harmonic generation (SHG) and
optical rectification (OR), as shown below.

\emph{i. Linear Stark effect.}---We first consider the Stark shift of quasiparticle frequencies induced by a static electric field.
For an applied dc field $E_{\rm dc}$, the
leading-order frequency shift can be evaluated
by the difference in the expectation values
of $H_{\rm ext}$ between the unperturbed
single-quasiparticle state $|k\rangle$
and the vacuum $|{\rm vac}\rangle$~\cite{notedipole},
\begin{equation}
\begin{aligned}
\hbar\Delta\omega_k
&\simeq
\langle k|H_{\rm ext}|k\rangle
-\langle{\rm vac}|H_{\rm ext}|{\rm vac}\rangle\\
&=-p_kE_{\rm dc}.
\end{aligned}
\label{eq:stark_shift}
\end{equation}
Since $p_k<0$, a positive electric field
$E_{\rm dc}>0$ aligned with the spontaneous
polarization produces a blueshift of the
quasiparticle excitation energy, whereas
reversing the field produces a redshift. The linear Stark shift thus provides
a direct spectroscopic signature of the
intrinsic electric dipole moment carried
by the quasiparticle.

\emph{ii. Coherent excitations and polarization response.}---An oscillating electric field, on the other hand,
can coherently excite the electronic dipolar modes and induce a dynamical
polarization response. We introduce the coherent amplitudes
$\beta_k=\langle b_k\rangle$ and the connected occupation and anomalous
pair correlators, $n_k=\langle b_k^\dagger b_k\rangle-|\beta_k|^2$ and $m_k=\langle b_kb_{-k}\rangle-\beta_k\beta_{-k}$, respectively. The quasiparticle-induced polarization response can then be expressed as~\cite{SM}
\begin{align}\delta P(t)=&-\frac{g(\beta_0+\beta_0^{\ast})}{\mathcal{V}}+\frac{1}{\mathcal V}\sum_k\left\{p_k|\beta_{k}|^2-\lambda_k(\beta_k\beta_{-k}\right.\nonumber\\&\left.+\beta_k^{\ast}\beta_{-k}^{\ast})\right\}+\frac{1}{\mathcal{V}}\sum_k[p_kn_k-\lambda_k(m_k+m_{-k}^\ast)],\label{eq:pol} \end{align}
with $\mathcal V$ the sample volume. The first two terms arise from coherent quasiparticle amplitudes,
whereas the last term accounts for the contributions
from connected quasiparticle correlations.

For a monochromatic electric field
$E(t)=E_\Omega e^{-i\Omega t}+\mathrm{c.c.}$,
with driving frequency $\Omega$ and complex
amplitude $E_\Omega$, we solve the equations
of motion for $\beta_k$, $m_k$, and $n_k$
perturbatively in the applied field, as detailed
in the Supplemental Material~\cite{SM}.
Substituting these solutions into Eq.~\eqref{eq:pol} leads to the linear polarization response $\delta P^{(1)}(t)=\epsilon_0
\chi^{(1)}E_\Omega e^{-i\Omega t}
+\mathrm{c.c.}$,
with the susceptibility
 \begin{align}
\chi^{(1)}=&
\frac{2}{\epsilon_0\mathcal V\hbar}\left\{
\frac{g^2\omega_{0}}{\omega_0^2-(\Omega+i\gamma)^2}+
\sum_k
\frac{4\lambda_k^2\omega_{k}}{(2\omega_k)^2-(\Omega+2i\gamma)^2}\right\},
\label{eq:linear_susceptibility}
\end{align}
where $\epsilon_0$ is the vacuum permittivity and $\gamma$ the phenomenological
single-quasiparticle damping rate. The first term describes the coherent excitation
of a single zero-momentum quasiparticle
and exhibits a resonance at $\Omega\simeq\omega_0$, whereas the second arises from the anomalous
pair correlator $m_k$, induced by the electric-field
coupling $\lambda_k b_k^\dagger b_{-k}^\dagger$ in Eq.~\eqref{eq:stark_bosonic}. This coupling directly creates
quasiparticle pairs, generating pair coherence at first order in the applied field
and giving rise to a two-quasiparticle
excitation continuum in the linear polarization
response at frequencies $\Omega\simeq 2\omega_k$. By contrast, the number-conserving quadratic coupling 
$\sim p_k b_k^\dagger b_k$ cannot excite quasiparticles
from the vacuum and hence does not contribute
to the linear susceptibility. The resulting polarization dynamics can emit
electromagnetic radiation and be probed by
terahertz pump--probe spectroscopy, as recently demonstrated for ferron
dynamics in van der Waals
ferroelectrics~\cite{Choe2026Ferron,Zhang2026Ferron}, offering a promising
experimental route to identifying purely
electronic dipolar collective excitations.


Beyond the linear response, electronic dipolar
collective excitations lead to second-harmonic generation (SHG) and optical
rectification (OR), enabled by the broken
inversion symmetry of the ordered state. Solving for $\beta_k$, $m_k$, and $n_k$
perturbatively to second order in the applied
electric field yields both a polarization
oscillating at $2\Omega$ and a rectified
dc polarization,
\begin{align}
\delta P^{(2)}(t)
=\epsilon_0\chi_{\rm SHG}^{(2)}
E_\Omega^2 e^{-2i\Omega t}
+\epsilon_0\chi_{\rm OR}^{(2)}
|E_\Omega|^2
+\mathrm{c.c.},
\label{eq:expansion}
\end{align}
where $\chi_{\rm SHG}^{(2)}(\Omega)$ and
$\chi_{\rm OR}^{(2)}(\Omega)$ characterize the
second-harmonic and optical rectification
responses, respectively, with their explicit
expressions provided in the Supplemental Material~\cite{SM}. As in the linear response, both the SHG and OR
susceptibilities contain coherent and
pair-correlation contributions.
The coherent contributions arise from
the interplay between the linear driving
term and the quadratic Stark couplings,
scaling as $g^2p_0$ and $g^2\lambda_0$.
By contrast, the pair-correlation
contributions scale as $p_k\lambda_k^2$
and involve quasiparticle pairs throughout
the Brillouin zone. 

\begin{figure}
    \centering
    \includegraphics[width=5.2 cm]{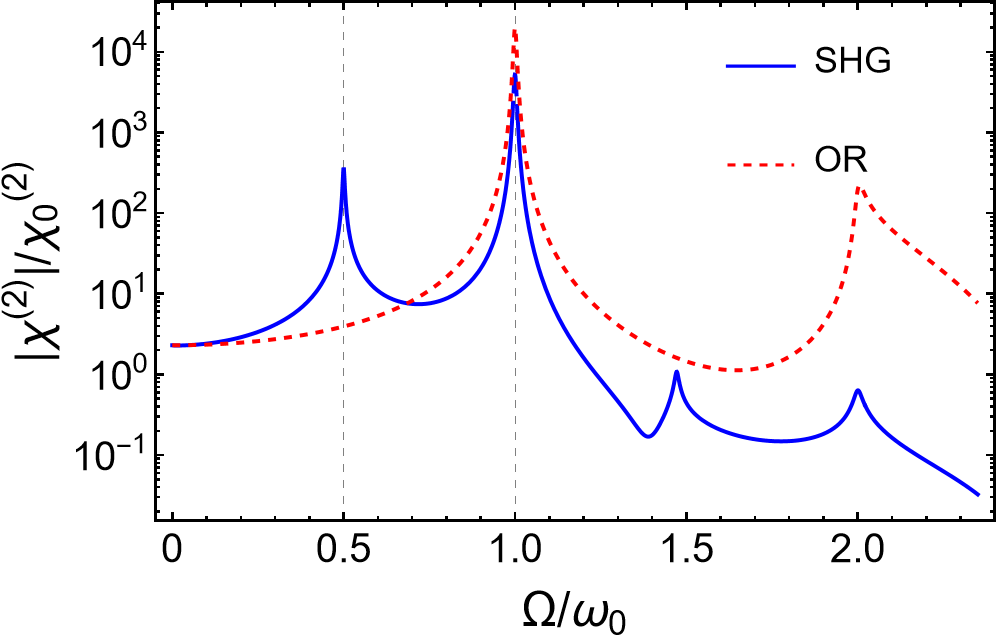}
    \caption{Second-harmonic generation (SHG) and optical rectification (OR) arising from dipolar quasiparticle excitations driven by an oscillating electric field at frequency $\Omega$.}
    \label{fig-2}
\end{figure}

Finally, we provide an order-of-magnitude estimate of the predicted properties and responses of dipolar quasiparticles. Given the limited knowledge of microscopic parameters in electronic ferroelectrics, we adopt representative parameters for the effective pseudospin-exchange Hamiltonian, $J_z=-30\,\mathrm{meV}$, $J_\perp=1\,\mathrm{meV}$, and $h_x=20\,\mathrm{meV}$, which stabilize a ferroelectrically ordered ground state. These parameters correspond to an extended Hubbard model with $V_0=100\,\mathrm{meV}$, $V_1=30\,\mathrm{meV}$, $t_0=10\,\mathrm{meV}$, $t_1=8\,\mathrm{meV}$, and $t_2=5\,\mathrm{meV}$. Taking an intradimer charge separation of $\delta=2\,\mathrm{\AA}$ and a lattice constant of $a=5\,\mathrm{\AA}$, we obtain a long-wavelength quasiparticle dispersion $\omega_k\simeq\sqrt{\omega_0^2+v_c^2k^2}$, with a gap frequency $\omega_0/2\pi\simeq5.7\,\mathrm{THz}$ and a characteristic velocity $v_c\simeq7.1\times10^3\,\mathrm{m/s}$; the intrinsic dipole moment of zone-center quasiparticles is $p_0\simeq-1.58\,e\mathrm{\AA}$, giving rise to a sizable Stark frequency shift of $\vert\Delta\omega_0\vert/2\pi\simeq38\,\mathrm{GHz}$ under a moderate static electric field of $10\,\mathrm{kV/cm}$. These characteristic quasiparticle properties are comparable in order of magnitude to those of ferrons in conventional ferroelectrics~\cite{Tang2022Ferron,Tang2024Ferron}.  

Fig.~\ref{fig-2} shows the magnitudes of $|\chi_{\rm SHG}^{(2)}|$ and $|\chi_{\rm OR}^{(2)}|$ as functions of the driving frequency $\Omega$, normalized by the characteristic susceptibility $\chi_{0}^{(2)}=g^2 \vert p_0\vert/[\epsilon_0\mathcal V(\hbar\omega_0)^2]$, for a quasiparticle quality factor $Q=\omega_0/(2\gamma)=100$. The SHG response exhibits two pronounced resonant peaks when either the fundamental frequency $\Omega$ or the second-harmonic frequency $2\Omega$ matches the zone-center quasiparticle frequency $\omega_0$, corresponding to one-photon and two-photon resonances, respectively. In contrast, the OR response exhibits a dominant resonance peak at $\Omega=\omega_0$, where the driving field directly excites the quasiparticle mode. Beyond these coherent resonances, quasiparticle-pair correlations give rise to additional, much weaker spectral features at higher frequencies, with enhanced responses near van Hove singularities in the two-quasiparticle density of states. Assuming an effective unit-cell volume of \(125\ {\rm \AA}^3\), we estimate \(\chi_{0}^{(2)}\simeq 2.2\times10^5\ {\rm pm/V}\) for the representative parameters considered here. The resulting second-order susceptibilities can reach values on the order of $100\,\mu\mathrm{m/V}$, several orders of magnitude larger than those in conventional nonlinear optical materials that originate from electronic interband transitions at much higher, optical (eV) energy scales~\cite{matlack2015review,abdelwahab2022giant,li2025giant}. Unlike nonlinear optical responses in structurally polar materials~\cite{boyd2019nonlinear},
the second-order responses identified here
originate from purely electronic dipolar
collective excitations without requiring
a polar lattice distortion, thereby providing
a complementary nonlinear spectroscopic probe
of electronic dipolar quasiparticles.

\emph{Conclusions.---}We derive an effective quantum dipolar-pseudospin
exchange model for electronic ferroelectricity in correlated
electron systems, revealing a close microscopic
analogy with magnetism. Just as spin exchange gives rise to magnons,
dipolar-pseudospin exchange supports collective
bosonic quasiparticles carrying intrinsic
electric dipole moments.
Unlike ferrons in conventional lattice-based ferroelectrics~\cite{Tang2022Ferron,bauer2022magnonics,Tang2024Ferron,tang2026multi}, these dipolar quasiparticles
originate from collective fluctuations of
the electronic charge distribution without
involving lattice vibrations. 
We show that their excitation spectrum can
be tuned by a static electric field through
a linear Stark effect, while an ac electric
field drives coherent excitations and induces
both linear and nonlinear polarization
responses, including exceptionally large, resonantly enhanced
second-harmonic generation and optical
rectification in the terahertz regime. These characteristic responses provide spectroscopic signatures of the dipolar quasiparticles and offer promising opportunities for electrically tunable nonlinear optics. Beyond optical spectroscopy, these quasiparticles may enable electric dipole transport in insulating electronic ferroelectrics, extending the paradigm of electronic transport beyond conventional charge currents and providing a purely electronic dipolar counterpart to magnon-mediated spin transport in magnetic insulators~\cite{Kajiwara2010,cornelissen2015long}.

\emph{Acknowledges.---}This work was supported by JSPS KAKENHI Grant-in-Aid for Scientific Research (B) (Grant No. 26K00625). 

\bibliography{reference}

@misc{notedipole,
    author = {},
    note ="{Here, we disregard the contribution arising from the static-field-induced reorientation of the classical ground-state pseudospins, which modifies the quasiparticle excitation frequency by affecting both the single-quasiparticle excited state and the vacuum state. This contribution is expected to be subdominant to the linear Stark shift associated with the intrinsic electric dipole moment of the quasiparticle.}"
}

@misc{SM,
    author = {},
    note ="{See the Supplemental Material for detailed derivations of the effective pseudospin exchange Hamiltonian Eq.~(\ref{eq:XXZ}) and the linear and nonlinear polarization responses of dipolar quasiparticles to alternating electric fields.}"
}

@article{de1963collective,
  title={Collective motions of hydrogen bonds},
  author={De Gennes, PG},
  journal={Solid State Communications},
  volume={1},
  number={6},
  pages={132--137},
  year={1963},
  doi={10.1016/0038-1098(63)90212-6},
  publisher={Elsevier}
}

@article{blinc1972dynamics,
  title={Dynamics of order-disorder-type ferroelectrics and anti-ferroelectrics},
  author={Blinc, Robert and {\v{Z}}ek{\v{s}}, Bo{\v{s}}tjan},
  journal={Advances in Physics},
  volume={21},
  number={93},
  pages={693--757},
  year={1972},
  url={https://doi.org/10.1080/00018737200101348},
  publisher={Taylor \& Francis}
}

@article{Imada1998,
  author  = {Imada, Masatoshi and Fujimori, Atsushi
             and Tokura, Yoshinori},
  title   = {Metal-insulator transitions},
  journal = {Rev. Mod. Phys.},
  volume  = {70},
  pages   = {1039--1263},
  year    = {1998},
  doi     = {10.1103/RevModPhys.70.1039}
}

@article{abdelwahab2022giant,
  title={Giant second-harmonic generation in ferroelectric NbOI2},
  author={Abdelwahab, Ibrahim and Tilmann, Benjamin and Wu, Yaze and Giovanni, David and Verzhbitskiy, Ivan and Zhu, Menglong and Bert{\'e}, Rodrigo and Xuan, Fengyuan and Menezes, Leonardo de S and Eda, Goki and others},
  journal={Nature Photonics},
  volume={16},
  number={9},
  pages={644--650},
  year={2022},
  url={https://doi.org/10.1038/s41566-022-01021-y},
  publisher={Nature Publishing Group UK, London}
}

@article{matlack2015review,
  title={Review of second harmonic generation measurement techniques for material state determination in metals},
  author={Matlack, Kathrn H and Kim, J-Y and Jacobs, Laurence J and Qu, Jianmin},
  journal={Journal of Nondestructive Evaluation},
  volume={34},
  number={1},
  pages={273},
  year={2015},
  url={https://doi.org/10.1007/s10921-014-0273-5},
  publisher={Springer}
}

@article{li2025giant,
  title={Giant second harmonic generation by photoinduced phase engineering},
  author={Li, Xinyue and Song, Boqin and Yan, Jiahao and Cao, Haijun and Xue, Tongtong and Chen, Jing and Dai, Yunyun and Du, Luojun and Zhang, Qinghua and Meng, Sheng and others},
  journal={Nano Letters},
  volume={25},
  number={14},
  pages={5821--5827},
  year={2025},
  url={https://doi.org/10.1021/acs.nanolett.5c00536},
  publisher={ACS Publications}
}

@article{Dagotto2005,
  author  = {Dagotto, Elbio},
  title   = {Complexity in Strongly Correlated
             Electronic Systems},
  journal = {Science},
  volume  = {309},
  pages   = {257--262},
  year    = {2005},
  doi     = {10.1126/science.1107559}
}

@article{Kajiwara2010,
  author  = {Kajiwara, Y. and Harii, K. and
             Takahashi, S. and Ohe, J. and
             Uchida, K. and Mizuguchi, M. and
             Umezawa, H. and Kawai, H. and
             Ando, K. and Takanashi, K. and
             Maekawa, S. and Saitoh, E.},
  title   = {Transmission of electrical signals
             by spin-wave interconversion
             in a magnetic insulator},
  journal = {Nature},
  volume  = {464},
  pages   = {262--266},
  year    = {2010},
  doi     = {10.1038/nature08876}
}

@article{Chumak2015,
  author  = {Chumak, A. V. and Vasyuchka, V. I.
             and Serga, A. A. and Hillebrands, B.},
  title   = {Magnon spintronics},
  journal = {Nat. Phys.},
  volume  = {11},
  pages   = {453--461},
  year    = {2015},
  doi     = {10.1038/nphys3347}
}

@article{cornelissen2015long,
  title={Long-distance transport of magnon spin information in a magnetic insulator at room temperature},
  author={Cornelissen, LJ and Liu, J and Duine, RA and Youssef, J Ben and Van Wees, BJ},
  journal={Nature Physics},
  volume={11},
  number={12},
  pages={1022--1026},
  year={2015},
  url={https://doi.org/10.1038/nphys3465},
  publisher={Nature Publishing Group UK, London}
}

@article{tang2026multi,
  title = {Multiferroic-like Quasiparticles in Ferroelectrics},
  author = {Tang, Ping and Bauer, Gerrit E. W.},
  journal = {Phys. Rev. Lett.},
  volume = {136},
  issue = {17},
  pages = {176702},
  numpages = {9},
  year = {2026},
  month = {May},
  publisher = {American Physical Society},
  doi = {10.1103/gm7n-8ftp},
  url = {https://link.aps.org/doi/10.1103/gm7n-8ftp}
}

@article{Adachi2023,
  title = {Ginzburg-Landau action and polarization current in an excitonic insulator model of electronic ferroelectricity},
  author = {Adachi, Hiroto and Ikeda, Naoshi and Saitoh, Eiji},
  journal = {Phys. Rev. B},
  volume = {107},
  issue = {15},
  pages = {155142},
  numpages = {10},
  year = {2023},
  month = {Apr},
  publisher = {American Physical Society},
  doi = {10.1103/PhysRevB.107.155142},
  url = {https://link.aps.org/doi/10.1103/PhysRevB.107.155142}
}

@article{PhysRevApplied.20.050501,
  title = {Polarization transport in ferroelectrics},
  author = {Bauer, G.E.W. and Tang, P. and Iguchi, R. and Xiao, J. and Shen, K. and Zhong, Z. and Yu, T. and Rezende, S.M. and Heremans, J.P. and Uchida, K.},
  journal = {Phys. Rev. Appl.},
  volume = {20},
  issue = {5},
  pages = {050501},
  numpages = {7},
  year = {2023},
  month = {Nov},
  publisher = {American Physical Society},
  doi = {10.1103/PhysRevApplied.20.050501},
  url = {https://link.aps.org/doi/10.1103/PhysRevApplied.20.050501}
}

@article{shen2025observation,
  title={Observation of ferron transport in ferroelectrics},
  author={Shen, Kaiwen and Tang, Ping and Chen, Xianzhe and Gao, Yifan and Fan, Yuanfei and Guo, Zejing and Wei, Yingfen and Jiang, Hao and Zhang, Xumeng and Wang, Ming and others},
  journal={arXiv preprint arXiv:2505.24419},
  url={https://doi.org/10.48550/arXiv.2505.24419},
  year={2025}
}

@article{itoh2026observation,
  title={Observation of nonlocal ferron-drag thermoelectricity},
  author={Itoh, Takuma and Hirai, Takamasa and Tang, Ping and Sepehri-Amin, Hossein and Iguchi, Ryo and Kozuka, Yusuke and Shimizu, Takao and Akita, Soshi and Mori, Shunsuke and Bauer, Gerrit EW and others},
  journal={arXiv preprint arXiv:2608.15596},
  url={https://doi.org/10.48550/arXiv.2608.15596},
  year={2026}
}

@article{PhysRevLett.126.187603,
  title = {Theory of Transport in Ferroelectric Capacitors},
  author = {Bauer, Gerrit E. W. and Iguchi, Ryo and Uchida, Ken-ichi},
  journal = {Phys. Rev. Lett.},
  volume = {126},
  issue = {18},
  pages = {187603},
  numpages = {4},
  year = {2021},
  month = {May},
  publisher = {American Physical Society},
  doi = {10.1103/PhysRevLett.126.187603},
  url = {https://link.aps.org/doi/10.1103/PhysRevLett.126.187603}
}

@article{PhysRevLett.128.047601,
  title = {Thermoelectric Polarization Transport in Ferroelectric Ballistic Point Contacts},
  author = {Tang, Ping and Iguchi, Ryo and Uchida, Ken-ichi and Bauer, Gerrit E. W.},
  journal = {Phys. Rev. Lett.},
  volume = {128},
  issue = {4},
  pages = {047601},
  numpages = {6},
  year = {2022},
  month = {Jan},
  publisher = {American Physical Society},
  doi = {10.1103/PhysRevLett.128.047601},
  url = {https://link.aps.org/doi/10.1103/PhysRevLett.128.047601}
}

@article{wooten2023electric,
  title={Electric field--dependent phonon spectrum and heat conduction in ferroelectrics},
  author={Wooten, Brandi L and Iguchi, Ryo and Tang, Ping and Kang, Joon Sang and Uchida, Ken-ichi and Bauer, Gerrit EW and Heremans, Joseph P},
  journal={Science advances},
  volume={9},
  number={5},
  pages={eadd7194},
  year={2023},
  doi={10.1126/sciadv.add7194},
  publisher={American Association for the Advancement of Science}
}

@article{tang2023nonlocal,
  title = {Nonlocal drag thermoelectricity generated by ferroelectric van der Waals heterostructures},
  author = {Tang, Ping and Uchida, Ken-ichi and Bauer, Gerrit E. W.},
  journal = {Phys. Rev. B},
  volume = {107},
  issue = {12},
  pages = {L121406},
  numpages = {6},
  year = {2023},
  month = {Mar},
  publisher = {American Physical Society},
  doi = {10.1103/PhysRevB.107.L121406},
  url = {https://link.aps.org/doi/10.1103/PhysRevB.107.L121406}
}

@article{jana2026ferron,
  title={Ferron-driven photoferroic hysteresis in van der Waals CuInP2S6},
  author={Jana, Sambhu and Zhang, Baolong and Mishra, Sobhan Subhra and Tan, Thomas CaiWei and Chaturvedi, Apoorva and Teo, Edwin Hang Tong and Singh, Ranjan},
  journal={Nature Communications},
  year={2026},
  url={https://doi.org/10.1038/s41467-026-73771-0},
  publisher={Nature Publishing Group UK, London}
}

@article{bauer2022magnonics,
  title={Magnonics vs. ferronics},
  author={Bauer, Gerrit EW and Tang, Ping and Iguchi, Ryo and Uchida, Ken-ichi},
  journal={Journal of Magnetism and Magnetic Materials},
  volume={541},
  pages={168468},
  year={2022},
  doi={10.1016/j.jmmm.2021.168468},
  publisher={Elsevier}
}

@article{kruglyak2010magnonics,
  title={Magnonics},
  author={Kruglyak, VV and Demokritov, SO and Grundler, D},
  journal={Journal of Physics D: Applied Physics},
  volume={43},
  number={26},
  pages={264001},
  doi={10.1088/0022-3727/43/26/264001},
  year={2010}
}

@article{PhysRev.127.965,
  title = {Broken Symmetries},
  author = {Goldstone, Jeffrey and Salam, Abdus and Weinberg, Steven},
  journal = {Phys. Rev.},
  volume = {127},
  issue = {3},
  pages = {965--970},
  numpages = {0},
  year = {1962},
  month = {Aug},
  publisher = {American Physical Society},
  doi = {10.1103/PhysRev.127.965},
  url = {https://link.aps.org/doi/10.1103/PhysRev.127.965}
}

@book{xu2013ferroelectric,
  title={Ferroelectric materials and their applications},
  author={Xu, Yuhuan},
  year={2013},
  publisher={Elsevier}
}

@article{Portengen1996,
  author  = {Portengen, T. and {\"O}streich, Th. and Sham, L. J.},
  title   = {Theory of electronic ferroelectricity},
  journal = {Phys. Rev. B},
  volume  = {54},
  pages   = {17452--17463},
  year    = {1996},
  doi     = {10.1103/PhysRevB.54.17452}
}

@article{vanDenBrink2008,
  author  = {van den Brink, J. and Khomskii, D. I.},
  title   = {Multiferroicity due to charge ordering},
  journal = {J. Phys.: Condens. Matter},
  volume  = {20},
  pages   = {434217},
  year    = {2008},
  doi     = {10.1088/0953-8984/20/43/434217}
}

@article{Hotta2010,
  author  = {Hotta, Chisa},
  title   = {Quantum electric dipoles in spin-liquid
             dimer Mott insulator
             {$\kappa$-(ET)$_2$Cu$_2$(CN)$_3$}},
  journal = {Phys. Rev. B},
  volume  = {82},
  pages   = {241104},
  year    = {2010},
  doi     = {10.1103/PhysRevB.82.241104}
}

@article{PhysRevLett.110.106401,
  title = {Collective Excitation of an Electric Dipole on a Molecular Dimer in an Organic Dimer-Mott Insulator},
  author = {Itoh, K. and Itoh, H. and Naka, M. and Saito, S. and Hosako, I. and Yoneyama, N. and Ishihara, S. and Sasaki, T. and Iwai, S.},
  journal = {Phys. Rev. Lett.},
  volume = {110},
  issue = {10},
  pages = {106401},
  numpages = {5},
  year = {2013},
  month = {Mar},
  publisher = {American Physical Society},
  doi = {10.1103/PhysRevLett.110.106401},
  url = {https://link.aps.org/doi/10.1103/PhysRevLett.110.106401}
}

@article{naka2012collective,
  title={Collective charge excitation in a dimer Mott insulating system},
  author={Naka, Makoto and Ishihara, Sumio},
  journal={journal of the physical society of japan},
  volume={82},
  number={2},
  pages={023701},
  year={2012},
  url={https://doi.org/10.7566/JPSJ.82.023701},
  publisher={The Physical Society of Japan}
}

@article{Monceau2001,
  author  = {Monceau, P. and Nad, F. Ya. and Brazovskii, S.},
  title   = {Ferroelectric Mott-Hubbard Phase of Organic
             (TMTTF)$_2$X Conductors},
  journal = {Phys. Rev. Lett.},
  volume  = {86},
  pages   = {4080--4083},
  year    = {2001},
  doi     = {10.1103/PhysRevLett.86.4080}
}

@article{Lunkenheimer2012,
  author  = {Lunkenheimer, P. and M{\"u}ller, J.
             and Krohns, S. and Schrettle, F.
             and Loidl, A. and Hartmann, B.
             and Rommel, R. and de Souza, M.
             and Hotta, C. and Schlueter, J. A.
             and Lang, M.},
  title   = {Multiferroicity in an Organic
             Charge-Transfer Salt That Is Suggestive
             of Electric-Dipole-Driven Magnetism},
  journal = {Nat. Mater.},
  volume  = {11},
  pages   = {755--758},
  year    = {2012},
  doi     = {10.1038/nmat3400}
}

@article{Yamauchi2009,
  title = {Ferroelectricity in multiferroic magnetite ${\text{Fe}}_{3}{\text{O}}_{4}$ driven by noncentrosymmetric ${\text{Fe}}^{2+}/{\text{Fe}}^{3+}$ charge-ordering: First-principles study},
  author = {Yamauchi, Kunihiko and Fukushima, Tetsuya and Picozzi, Silvia},
  journal = {Phys. Rev. B},
  volume = {79},
  issue = {21},
  pages = {212404},
  numpages = {4},
  year = {2009},
  month = {Jun},
  publisher = {American Physical Society},
  doi = {10.1103/PhysRevB.79.212404},
  url = {https://link.aps.org/doi/10.1103/PhysRevB.79.212404}
}

@article{ikeda2005ferroelectricity,
  title={Ferroelectricity from iron valence ordering in the charge-frustrated system LuFe2O4},
  author={Ikeda, Naoshi and Ohsumi, Hiroyuki and Ohwada, Kenji and Ishii, Kenji and Inami, Toshiya and Kakurai, Kazuhisa and Murakami, Youichi and Yoshii, Kenji and Mori, Shigeo and Horibe, Yoichi and others},
  journal={Nature},
  volume={436},
  number={7054},
  pages={1136--1138},
  year={2005},
  url={https://doi.org/10.1038/nature04039},
  publisher={Nature Publishing Group UK, London}
}

@article{ishihara2014electronic,
  title={Electronic ferroelectricity in molecular organic crystals},
  author={Ishihara, Sumio},
  journal={Journal of Physics: Condensed Matter},
  volume={26},
  number={49},
  pages={493201},
  year={2014},
  doi={10.1088/0953-8984/26/49/493201},
  publisher={IOP Publishing}
}

@article{naka2010electronic,
  title={Electronic ferroelectricity in a dimer Mott insulator},
  author={Naka, Makoto and Ishihara, Sumio},
  journal={Journal of the Physical Society of Japan},
  volume={79},
  number={6},
  pages={063707},
  year={2010},
  url={https://doi.org/10.1143/JPSJ.79.063707},
  publisher={The Physical Society of Japan}
}

@article{PhysRevLett.85.1698,
  title = {Charge Ordering in the TMTTF Family of Molecular Conductors},
  author = {Chow, D. S. and Zamborszky, F. and Alavi, B. and Tantillo, D. J. and Baur, A. and Merlic, C. A. and Brown, S. E.},
  journal = {Phys. Rev. Lett.},
  volume = {85},
  issue = {8},
  pages = {1698--1701},
  numpages = {0},
  year = {2000},
  month = {Aug},
  publisher = {American Physical Society},
  doi = {10.1103/PhysRevLett.85.1698},
  url = {https://link.aps.org/doi/10.1103/PhysRevLett.85.1698}
}

@article{Giovannetti2015,
  author  = {Giovannetti, Gianluca
             and Nourafkan, Reza
             and Kotliar, Gabriel
             and Capone, Massimo},
  title   = {Correlation-driven electronic
             multiferroicity in organic crystals},
  journal = {Phys. Rev. B},
  volume  = {91},
  pages   = {125130},
  year    = {2015},
  doi     = {10.1103/PhysRevB.91.125130}
}

@article{tang2026pseudospin,
  title={Bichargon Hall Effect},
  author={Tang, Ping},
  journal={arXiv preprin arXiv:2610.07719},
  url={https://doi.org/10.48550/arXiv.2610.07719},
  year={2026}
}

@article{tang2026collective,
  title={Collective Charge-$\backslash$(2e$\backslash$) Bosonic Excitations in Charge-Ordered Systems},
  author={Tang, Ping},
  journal={arXiv preprint arXiv:2609.20472},
  year={2026},
  url={
https://doi.org/10.48550/arXiv.2609.20472},
}

@book{giamarchi2004quantum,
  title={Quantum physics in one dimension},
  author={Giamarchi, Thierry and others},
  volume={121},
  year={2004},
  publisher={Clarendon, Oxford}
}

@article{Takahashi1977,
  author  = {Takahashi, Minoru},
  title   = {Half-filled Hubbard model at low temperature},
  journal = {J. Phys. C: Solid State Phys.},
  volume  = {10},
  pages   = {1289--1301},
  year    = {1977},
  doi     = {10.1088/0022-3719/10/8/031}
}

@book{rezende2020fundamentals,
  title={Fundamentals of magnonics},
  author={Rezende, Sergio M},
  volume={969},
  year={2020},
  publisher={Springer}
}

@article{Ising1925,
  author  = {Ising, E.},
  title   = {Beitrag zur Theorie des Ferromagnetismus},
  journal = {Zeitschrift f{\"u}r Physik},
  volume  = {31},
  pages   = {253--258},
  year    = {1925},
  doi     = {10.1007/BF02980577}
}

@article{rezende2019introduction,
  title={Introduction to antiferromagnetic magnons},
  author={Rezende, Sergio M and Azevedo, Antonio and Rodr{\'\i}guez-Su{\'a}rez, Roberto L},
  journal={Journal of Applied Physics},
  volume={126},
  number={15},
  year={2019},
  url={https://doi.org/10.1063/1.5109132},
  publisher={AIP Publishing}
}

@incollection{boyd2019nonlinear,
  title={Nonlinear optics},
  author={Boyd, Robert W},
  booktitle={Handbook of Laser Technology and Applications (Three-Volume Set)},
  pages={180--202},
  year={2019},
  publisher={CRC Press}
}

@article{Scalapino1975,
  author  = {Scalapino, D. J. and Imry, Y. and Pincus, P.},
  title   = {Generalized Ginzburg-Landau theory of
             pseudo-one-dimensional systems},
  journal = {Phys. Rev. B},
  volume  = {11},
  pages   = {2042--2048},
  year    = {1975},
  doi     = {10.1103/PhysRevB.11.2042}
}

@article{tang2026bichargon,
  title={Bichargon Hall Effect},
  author={Tang, Ping},
  url={https://doi.org/10.48550/arXiv.2610.0771},
  journal={arXiv preprint arXiv:2610.07719},
  year={2026}
}

@article{MacDonald1988,
  author  = {MacDonald, A. H. and Girvin, S. M.
             and Yoshioka, D.},
  title   = {{$t/U$} expansion for the Hubbard model},
  journal = {Phys. Rev. B},
  volume  = {37},
  pages   = {9753--9756},
  year    = {1988},
  doi     = {10.1103/PhysRevB.37.9753}
}

@article{Anderson1959,
  author  = {Anderson, P. W.},
  title   = {New Approach to the Theory of
             Superexchange Interactions},
  journal = {Phys. Rev.},
  volume  = {115},
  pages   = {2--13},
  year    = {1959},
  doi     = {10.1103/PhysRev.115.2}
}

@article{Ishihara2000,
  author  = {Sumio Ishihara and Sadamichi Maekawa},
  title   = {Theory of orbital excitation and resonant
             inelastic x-ray scattering in manganites},
  journal = {Phys. Rev. B},
  volume  = {62},
  pages   = {2338},
  year    = {2000},
  doi     = {10.1103/PhysRevB.62.2338}
}

@article{Saitoh2001,
  author  = {E. Saitoh and S. Okamoto and K. T. Takahashi
             and K. Tobe and K. Yamamoto and T. Kimura
             and S. Ishihara and S. Maekawa and Y. Tokura},
  title   = {Observation of orbital waves as elementary
             excitations in a solid},
  journal = {Nature},
  volume  = {410},
  pages   = {180--183},
  year    = {2001},
  doi     = {10.1038/35065547}
}

@article{vanDenBrink2001,
  author  = {Jeroen van den Brink},
  title   = {Orbital Excitations in {LaMnO$_3$}},
  journal = {Phys. Rev. Lett.},
  volume  = {87},
  pages   = {217202},
  year    = {2001},
  doi     = {10.1103/PhysRevLett.87.217202}
}

@article{Schlappa2012,
  author  = {J. Schlappa and K. Wohlfeld and K. J. Zhou
             and M. Mourigal and M. W. Haverkort
             and V. N. Strocov and L. Hozoi and C. Monney
             and S. Nishimoto and S. Singh
             and A. Revcolevschi and J.-S. Caux
             and L. Patthey and H. M. R{\o}nnow
             and J. van den Brink and T. Schmitt},
  title   = {Spin-orbital separation in the
             quasi-one-dimensional Mott insulator
             {Sr$_2$CuO$_3$}},
  journal = {Nature},
  volume  = {485},
  pages   = {82--85},
  year    = {2012},
  doi     = {10.1038/nature10974}
}

@incollection{spaldin2007analogies,
  title={Analogies and differences between ferroelectrics and ferromagnets},
  author={Spaldin, Nicola A},
  booktitle={Physics of Ferroelectrics: A Modern Perspective},
  pages={175--218},
  year={2007},
  url={https://link.springer.com/book/10.1007/978-3-540-34591-6#page=188},
  publisher={Springer}
}

@article{Cochran1960,
  author  = {W. Cochran},
  title   = {Crystal stability and the theory of ferroelectricity},
  journal = {Adv. Phys.},
  volume  = {9},
  pages   = {387--423},
  year    = {1960},
  doi     = {10.1080/00018736000101229}
}

@article{Tang2022Ferron,
  author  = {Ping Tang and Ryo Iguchi and Ken-ichi Uchida
             and Gerrit E. W. Bauer},
  title   = {Excitations of the ferroelectric order},
  journal = {Phys. Rev. B},
  volume  = {106},
  pages   = {L081105},
  year    = {2022},
  doi     = {10.1103/PhysRevB.106.L081105}
}

@article{Tang2024Ferron,
  author  = {Ping Tang and Gerrit E. W. Bauer},
  title   = {Electric analog of magnons in order-disorder
             ferroelectrics},
  journal = {Phys. Rev. B},
  volume  = {109},
  pages   = {L060301},
  year    = {2024},
  doi     = {10.1103/PhysRevB.109.L060301}
}

@article{Choe2026Ferron,
  author  = {Jeongheon Choe and Taketo Handa
             and Chun-Ying Huang and others},
  title   = {Observation of coherent ferron emission
             and propagation},
  journal = {Nat. Mater.},
  year    = {2026},
  doi     = {10.1038/s41563-026-02597-4}
}

@article{Zhang2026Ferron,
  author  = {Baolong Zhang and Ruihuan Duan
             and Sobhan Subhra Mishra and Sambhu Jana
             and Jonghyeon Kim and others},
  title   = {Electric-field control of coherent
             ferron oscillations},
  journal = {Nat. Phys.},
  year    = {2026},
  doi     = {10.1038/s41567-026-03406-0}
}

@article{Cochran1959,
  author  = {W. Cochran},
  title   = {Crystal Stability and the Theory of Ferroelectricity},
  journal = {Phys. Rev. Lett.},
  volume  = {3},
  pages   = {412--414},
  year    = {1959},
  doi     = {10.1103/PhysRevLett.3.412}
}
\end{document}